\documentclass[12pt,preprint]{emulateapj}
\usepackage{color}
\shorttitle{Environmental Effects on Galaxy Clustering}
\shortauthors{Zu et al.}
\def\xizr{\xi_{0/R}}

\def\xiQp{Q_\xi}
\def\rhalf{r_{\xi/2}}
\def\hMsun{h^{-1}M_\odot}
\def\hMpc{h^{-1}{\rm Mpc}}
\def\Mvir{M_{\rm vir}}
\def\rvec{{\bf r}}
\def\xvec{{\bf x}}
\def\la{\langle}
\def\ra{\rangle}

\begin{document}
\title{Environmental Effects on Real-Space and Redshift-Space Galaxy Clustering}
\author{ 
Ying Zu\altaffilmark{1,2},
Zheng Zheng\altaffilmark{3,4,5}, 
Guangtun Zhu\altaffilmark{1,6}, 
and Y.P. Jing\altaffilmark{1} 
} 

\altaffiltext{1}{Shanghai Astronomical Observatory, Joint Institute for Galaxy and Cosmology
  (JOINGC) of SHAO and USTC, Nandan Road 80, Shanghai, 200030, China; nye@shao.ac.cn ;
ypjing@shao.ac.cn.}
\altaffiltext{2}{Graduate School of the Chinese Academy of Sciences, 19A, Yuquan Road, Beijing, China}
\altaffiltext{3}{Institute for Advanced Study, Einstein Drive,
  Princeton, NJ 08540; zhengz@ias.edu. } 
\altaffiltext{4}{Hubble Fellow.}
\altaffiltext{5}{John Bahcall Fellow.}
\altaffiltext{6}{Center
  for Cosmology and Particle Physics, New York University, New York, NY10003, USA; gz323@nyu.edu.
}

\begin{abstract}

  Galaxy formation inside dark matter halos, as well as the halo formation itself, can be affected
  by large-scale environments. Evaluating the imprints of environmental effects on galaxy
  clustering is crucial for precise cosmological constraints with data from galaxy redshift
  surveys. We investigate such an environmental impact on both real-space and redshift-space
  galaxy clustering statistics using a semi-analytic model (SAM) derived from the Millennium
  Simulation. We compare clustering statistics from original SAM galaxy samples and shuffled ones
  with environmental influence on galaxy properties eliminated. Among the luminosity-threshold
  samples examined, the one with the lowest threshold luminosity ($\sim 0.2L_*$) is affected by
  environmental effects the most, which has a $\sim$10\% decrease in the real-space two-point
  correlation function~(2PCF) after shuffling. By decomposing the 2PCF into five different
  components based on the source of pairs, we show that the change in the 2PCF can be explained by
  the age and richness (galaxy occupation number) dependence of halo clustering. The 2PCFs in
  redshift space are found to change in a similar manner after shuffling. If the environmental
  effects are neglected, halo occupation distribution modeling of the real-space and
  redshift-space clustering may have a less than 6.5\% systematic uncertainty in constraining
  $\sigma_{8}\Omega_{m}^{0.6}$ from the most affected SAM sample and have substantially smaller
  uncertainties from the other, more luminous samples. We argue that the effect could be even
  smaller in reality. In the Appendix, we present a method to decompose the 2PCF, which can be
  applied to measure the two-point auto-correlation functions of galaxy sub-samples in a
  volume-limited galaxy sample and their two-point cross-correlation functions in a single run
  utilizing only one random catalog.

\end{abstract}


\section{Introduction}
 
Recently, many authors have identified the environmental impact, which manifests itself as another
degree of freedom, on the clustering of halos at fixed mass. \cite{GSW05} found that low mass
halos ($M<M_{\ast}$) which formed earlier are more strongly clustered than their younger
counterparts; whilst for high mass halos, older halos with $M>10M_{\ast}$ turn out to be less
clustered than the younger ones (\citealt{Wechsler06,JSM07,Wetzel07}), where $M_{\ast}$ is the
nonlinear mass scale for collapse. This environmental dependence of halo clustering, namely
``assembly bias'' (\citealt{GW07}), contradicts with excursion set theory (EST;
\citealt{Bond91,LC93,MW96}) which predicts that an individual halo evolves without awareness of
the larger environment except for its own mass when it is {\it observed} \citep{White96}. In this
paper, we investigate the effect of the halo assembly bias on modeling the real-space and
redshift-space galaxy clustering statistics and discuss the possible consequence on cosmological
parameter constraints from these clustering statistics.

Several possible explanations are proposed to decode halo assembly bias by either studying
detailed halo growth within high-resolution $N$-body simulations or by improving current excursion
set theory. \cite{Wang07} showed that the accretion of low mass halos in dense regions is severely
truncated due to tidal disruption and preheating by their massive companions, and \cite{JSM07}
suggested that the competition for accretion resources also triggers a delayed accretion phase
which results in the inverse age-dependence of massive halos, while \citet{Keselman07} argued that
highly non-linear effects like tidal stripping may not be the main driver for assembly bias. On
the other hand, \cite{Zetner06} implemented a toy model by substituting the sharp-$k$ filter in
EST with a localized configuration one, and \cite{Sandvik07} integrated EST with ellipsoidal
collapse model and barrier-crossing of pancakes and filaments. Both theoretical trials claimed
that the assembly bias for massive halos could be naturally recovered, at least partly offset, by
deserting Markovian simplification in EST. Recently, \citet{Dalal08} showed that the assembly bias
of rare massive halos is expected from the statistics of peaks in Gaussian random fields, and they
argued that the formation of a non-accreting sub-population of low-mass halos is responsible for
the assembly bias of low mass halos (also see \citealt{Hahn08}).

As the products of gas physics within dark matter halos, galaxies have no reason to be immune from
this environmental effect. \cite{Croton07} and \cite{Zhu06} showed that environmental effect is
transmitted to the clustering and properties of galaxies in semi-analytic model (SAM) and smoothed
particle hydrodynamics~(SPH) simulation, both of which extract halo merging histories directly
from simulations rather than Markovian process. Observationally, \cite{YMB06} and \cite{Berlind06}
found a residual dependence of galaxy clustering on group properties other than group mass by
using group catalog from the Two-Degree Field Galaxy Redshift Survey (2dFGRS; \citealt{Colless01})
and the Sloan Digital Sky Survey (SDSS; \citealt{York00}), respectively.

In modeling the galaxy clustering, the halo occupation distribution~(HOD) or the closely related
conditional luminosity function (CLF) is a powerful method to put the observed galaxy clustering
in an informative form of describing the relation between galaxies and dark matter halos
(\citealt{JMB98,Seljak00,PS00,Scoccimarro01,CS02,BW02,Yang03,Zheng05}). It successfully explains
the departure from a power-law in the observed galaxy two-point correlation functions
(\citealt{Zehavi05}) and bridges the gap between high-resolution $N$-body simulations of dark
matter particles and large scale surveys of galaxies. HOD modeling also enhances the power of
galaxy clustering on constraining cosmological parameters by linking galaxies to dark matter halos
and using the clustering data on all scales (\citealt{Bosch03,Abazajian05,Zheng07}). However, one
key assumption in the current version of the HOD is based on the EST that the formation and the
distribution of galaxies within halos are {\it statistically} determined solely by halo mass.
Therefore, it is important to quantify the environmental effect on modeling galaxy clustering
within the HOD framework in the era of precision cosmology and provide insights to improve the HOD
modeling.

A natural way to study the environmental effect is to extend the current HOD framework by
including a second halo variable other than halo mass and compare the modeling results with
previous results. Many candidates of halo variables, such as formation time, concentration,
substructure richness, and spin, have been scrutinized but all were proved incapable of capturing
environmental effect neatly and completely (\citealt{GW07,Wechsler06}). One of the reasons for
this is that halo formation history is subject to incidental merging events and uneven accretion
phases, both of which produce a large scatter in the relation between any halo property and the
environment (\citealt{Wechsler02,Zhao03}).

In the present study, we shuffle a semi-analytic galaxy sample to produce three sets of artificial
samples, which either partly or completely lost their environmental features, and investigate the
changes in real-space and redshift-space galaxy clustering statistics. The shuffling would enable
us to see the consequences of neglecting the environmental dependence in the current version of
HOD modeling and give us ideas of the effect on constraining cosmological parameters using these
statistics (e.g., \citealt{Tinker06}). The structure of the paper is as follows. In \S~2, we
introduce the simulation and the SAM model we use and describe our construction of galaxy samples
with different threshold luminosities from the SAM. In \S~3, we present three methods of shuffling
the galaxy samples aimed to eliminate the environmental dependence. Then, in \S~4, we analyze in
detail the effect of environments on the real-space two-point correlation functions (2PCFs) by
comparing the results between samples before and after shuffling. In \S~5, we study the effect of
environments on the redshift-space clustering statistics. We conclude in \S~6 with a brief
discussion and summary. In the Appendix, we present a method to decompose the 2PCFs into different
components based on the properties of galaxy pairs. This method can be generalized to apply to
real data to measure the two-point auto-correlation functions of galaxy sub-samples in a
volume-limited galaxy sample and their two-point cross-correlation functions in a single run
utilizing only one random catalog.
 
\section{Simulation Data and Semi-Analytic Model}

In this study, we make use of outputs from a galaxy formation model based on the Millennium
Simulation. The Millennium Simulation \citep{Springel05} follows the hierarchical growth of dark
matter structures from redshift $z=127$ to the present. The simulation adopts a concordance
cosmological model with ($\Omega_{m}$, $\Omega_{\Lambda}$, $\Omega_{b}$, $\sigma_{8}$, $h$)=(0.25,
0.75, 0.045, 0.9, 0.73), and employs 2160$^{3}$ particles of mass $8.6\times 10^{8} \hMsun$ in a
periodic box with comoving size 500~$\hMpc$ on a side. Friends-Of-Friends (FOF; \citealt{Davis85})
halos are identified in the simulation at each of the $64$ snapshots with a linking length 0.2
times the mean particle separation. Substructures are then identified by SUBFIND algorithm as
locally overdense regions in the background FOF halos \citep{Springel01}. Detailed merger trees of
all gravitationally self-bound dark matter clumps constructed from this simulation provide a key
ingredient for semi-analytic models of galaxy formation.

The galaxy catalog we use is from the semi-analytic model (SAM) of \cite{DB07}, which is an
updated version of that of \cite{Croton06} and \cite{DeLucia06}. This model explores a variety of
physical processes related to galaxy formation. It can reproduce many observed properties of
galaxies in the local universe, including the galaxy luminosity function, the bimodal distribution
of colors, the Tully-Fisher relation, the morphology distribution, and the 2PCFs for various type
and luminosity selected samples. This particular model is of course not guaranteed to be
absolutely right. What is important to our study here is that the environment-dependent
ingredients inherent in this model, such as the history of dynamical interactions and mergers of
halos, should be well transmitted to and preserved in the resultant galaxy population. We aim to
investigate the likely effects of the environmental dependence in this model on galaxy clustering
statistics in real space and redshift space and explore the implications for cosmological study
with galaxy clustering data.

We construct six luminosity-threshold galaxy samples at $z=0$ from the SAM catalog according to
the restframe SDSS $r$-band absolute magnitude $M_{r}$ with dust extinction included. 
Table~\ref{tab:tab1} lists the properties of these samples. Our L207 sample has a number density 
similar to the observed $L>L_{*}$ sample (see Table~2 of \citealt{Zehavi05}). Since more luminous 
galaxies tend to reside in more massive halos (e.g., \citealt{Zehavi05}), these six samples can 
probe different halo mass ranges (from mass below $M_*$ to that above $M_*$). The halo assembly 
bias has different amplitudes and signs across these mass ranges (e.g., \citealt{GSW05,GW07, 
Wechsler06,JSM07}), we therefore expect different environmental effects from the six samples.

\begin{deluxetable}{cccrr}
\tablewidth{0pt}
\tablecolumns{5}
\tablecaption{\label{tab:tab1}
Properties of the Luminosity-threshold Samples
}
\tablehead{
Name & $M_{r}^{\rm max}$ & $\bar{n}$ ($10^{-2}h^3{\rm Mpc}^{-3}$) & N$_{\rm gal}$ & N$_{\rm halo}$
}
\startdata
L190 &  -19.0 & 1.835  & 2293947  &  1661007\\
L200 &  -20.0 & 0.771  &  963452  &   730330\\
L207 &  -20.7 & 0.293  &  365845  &   289226\\
L210 &  -21.0 & 0.167  &  209206  &   169864\\
L217 &  -21.7 & 0.032  &  39402   &    34129\\
L220 &  -22.0 & 0.013  &  16084   &    14148\\
\enddata
\end{deluxetable}

\section{Shuffling Schemes}

Our purpose in this paper is to study the impact of environmental dependence on the HOD modeling
of real-space and redshift-space clustering statistics. Besides the six galaxy samples from the
SAM, for comparison we also need galaxy samples with the environmental dependence eliminated.
Following \cite{Croton07}, we construct such control samples from the original SAM catalog by
shuffling galaxy contents in halos of similar masses. We produce three sets of control samples
based on three shuffling schemes described below.

We first group all the FOF dark matter halos at $z=0$ with $\Mvir$ larger than $5.5 \times
10^{10}\hMsun$ in the catalog into different mass bins of width $\Delta\log[\Mvir/(\hMsun)]=0.1$.
Then we record the relative positions and velocities of all the satellites to their affiliated
central galaxies, whose positions and velocities are set to those of their host halos in the SAM.
Finally, we redistribute the galaxies within individual halo mass bins. The three sets of our
control samples (hereafter CTL1, CTL2 and CTL3, respectively) differ in the way how galaxies are
re-distributed.

For CTL1, we follow the scheme of \cite{Croton07} to keep the original configuration of galaxies
inside each halo intact and move the galaxy content to its new host as a whole. In this way, the
one-halo term contribution to the galaxy clustering statistics is almost unchanged. In order to
compensate for the non-zero mass bin effect in the shuffling, we scale the recorded relative
position and velocity of each galaxy by $(M_{\rm new}/M_{\rm old})^{1/3}$ in order to redistribute
the galaxies in the original halo of mass $M_{\rm old}$ to the new host halo of mass $M_{\rm
  new}$. This improvement ensures that the position of shuffled galaxy content be regulated by the
virial radius of new host halos.

For CTL2, we collect the distance $r$ to the halo center and the velocity $v$ relative to the halo
center for all the satellite galaxies that belong to halos in the same mass bin. Then a pair of
$r$ and $v$ are randomly drawn from the sets and assigned to a galaxy. This galaxy is put into a
randomly selected halo in that mass bin with random orientations for both ${\bf r}$ and ${\bf v}$
[with the $(M_{\rm new}/M_{\rm old})^{1/3}$ scaling applied]. For central galaxies, they are
randomly assigned to halos of the same mass bin. This shuffling procedure assumes a mean radial
galaxy number density profile for all halos in the same mass bin and completely eliminates any
environmental features in the galaxy distribution inside halos, including the alignment and
segregation of satellites, the non-spherical shape of halos, the infall pattern of satellite
velocity distribution, and any correlation between central and satellite galaxies (e.g., in
luminosity). CTL2 would allow us to infer the largest effect that environment may have on galaxy
clustering statistics for the given SAM.

In addition to CTL1 and CTL2, we construct another set of samples (CTL3) by isotropizing satellites
inside their own halo without shuffling contents between different halos. In this way, the radial
distribution of galaxies in each individual halo is conserved, but the statistical angular 
distribution loses the anisotropy. CTL3 allows us to isolate the effect of assuming spherical
symmetry for the satellite distribution in modeling 2PCFs. Although CTL2 also isotropizes the
satellite distribution inside halos, it effectively uses a radial distribution averaged over
halos of similar masses. Therefore, comparing CTL3 and CTL2 would show the effect of the 
scatter in the distributions of satellites in halos of similar masses.

For each of the CTL1, CTL2, and CTL3 shuffling schemes, we create 10 different galaxy catalogs 
varying the random seed. We extract the 6 control luminosity-threshold samples from each 
shuffled catalog in accordance with the above L190, L200, L207, L210, L217 and L220 samples of 
the SAM. To prevent
numerical effects from mixing with the physical effects we are to ascertain, we have performed
tests by reducing the size of halo mass bins or leaving several most massive bins unshuffled and
find that our choice of the bin size does not introduce noticeable numerical effect.

\begin{figure}
  \epsscale{1.1} \plotone{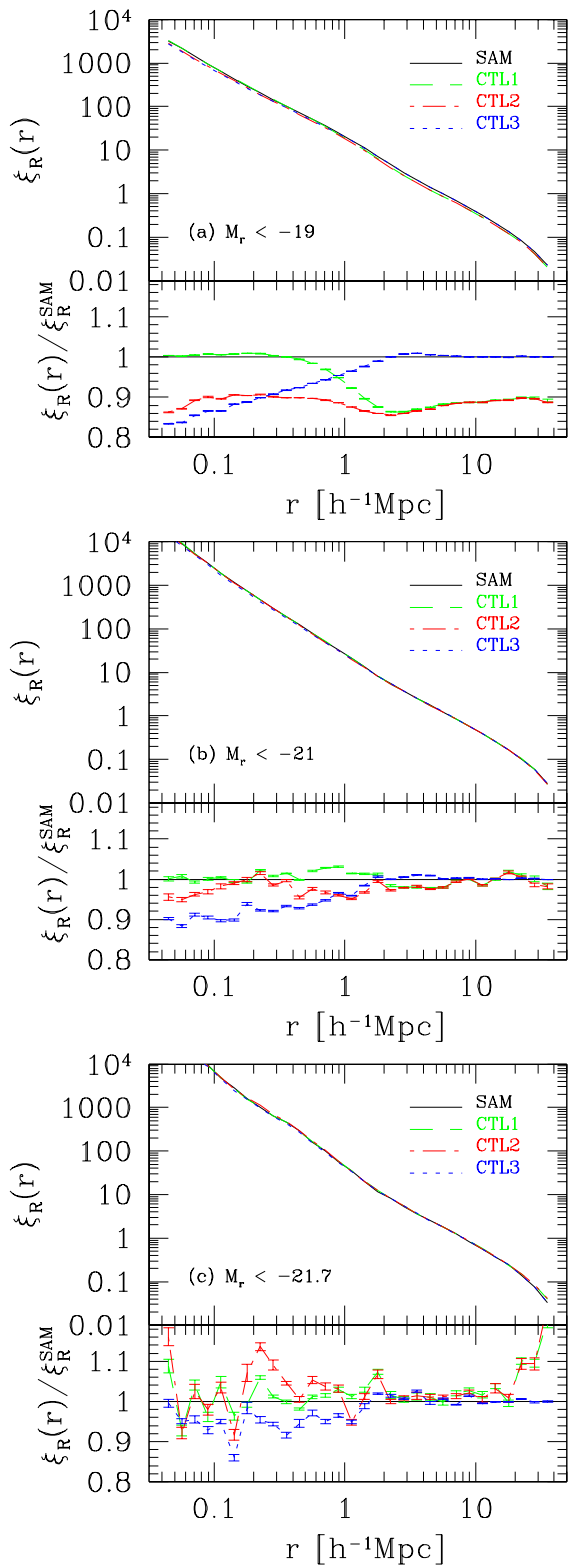}
  \caption{The comparison of 2PCFs between SAM and shuffled samples for three different luminosity
    thresholds. The lower part of each panel gives the ratio of the 2PCFs of the shuffled and the
    original SAM samples. Solid lines are the 2PCFs for SAM samples, while dashed, dot-dashed, and
    dotted lines are those for the CTL1, the CTL2 and the CTL3 shuffled samples, respectively. 
    See the text. The (small) error bars reflect the scatter from the 10 realizations for each 
    shuffled sample. }
  \label{bias}
\end{figure}

\section{Environmental Effect on Real-Space 2PCFs}
\label{sec:realspace}

We start from comparing the real-space 2PCFs of the original SAM samples and the shuffled samples. 
The 2PCFs essentially describe the pair count as a function of pair separation. On small (large)
scales, galaxy pairs are dominated by one-halo (two-halo) pairs, i.e., intra-halo (inter-halo) pairs.

\begin{figure}
  \epsscale{1.2} \plotone{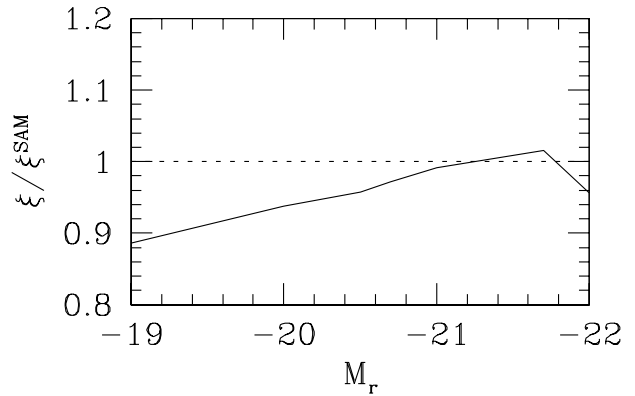}
  \caption{The ratio of the 2PCF of shuffled sample to that of the SAM sample as a function of
    magnitude limit of sample, averaged over scales of 5--25$\hMpc$. In the SAM catalog used in
    this paper, $L_*$ corresponds to $M_r\sim -20.7$. Since the ratio is nearly 
    constant on those scales, we do not show the small error bars.}
  \label{b2}
\end{figure}

Our results on the real-space 2PCFs are shown in Figure~\ref{bias}. On small scales ($\lesssim
2\hMpc$), where the one-halo term dominates, 2PCFs from CTL1, CTL2 and CTL3 behave differently. In
CTL1, galaxy contents inside halos as a whole are exchanged among halos of similar mass, so we do
not expect any appreciable change in the one-halo regime of the 2PCFs. Thus, the small-scale
clustering in CTL1 remains almost the same as that of the original one, as seen in
Figure~\ref{bias}. The slight differences seen in the plot are a result of the finite mass bin.
CTL3 makes satellite distribution inside halos isotropic, which on average enlarge the separations
of intra-halo galaxy pairs. So on small scales, 2PCFs of CTL3 are always smaller than those of SAM
with a suppression of around 10\%. In CTL2, not only the angular distribution of satellite
galaxies inside halos are isotropized but the radial galaxy number density profile is averaged
within the same mass bin, which completely erases the memories of galaxies about their
environments. Figure~\ref{bias}$a$ shows that galaxies in CTL2 exhibit a suppression up to
$\sim$10\% for L190 samples and the suppression becomes weaker for samples with higher threshold
luminosity (e.g., sample L210 in Fig.~\ref{bias}$b$). The 2PCF for the L217 CTL2 sample
(Fig.~\ref{bias}$c$) is too noisy to tell the trend, but it is likely to still be a suppression
(see below).

On large scales, where the two-halo term dominates, 2PCFs from CTL3 stay the same as those from
the original SAM samples, since the CTL3 scheme only shuffles galaxies in each halo. It is not
surprising that the large-scale 2PCFs of CTL1 and CTL2 are almost identical, given that they both
shuffle halos of similar mass. The difference between the 2PCF in the CTL1/CTL2 sample and the
original SAM sample shows a steady trend with the threshold luminosity. For the faint sample
($M_r<-19$), the environmental dependence of the galaxy population and the assembly bias lead to a
$\sim$10\% suppression in the 2PCF after shuffling (Fig.~\ref{bias}$a$). For the intermediate
sample $M_r<-21$, the difference between shuffled and SAM samples is reduced to $\sim$2\%
(Fig.~\ref{bias}$b$). For the bright $M_{r}<-21.7$ sample, the 2PCFs of shuffled ones become
$\sim$3\% larger than those of the SAM sample (Fig.~\ref{bias}$c$). To see the trend more clearly,
we show in Figure~\ref{b2} the ratio $\xi/\xi^{\rm SAM}$ of large-scale 2PCFs of the shuffled
(CTL1 or CTL2) and the SAM samples as a function of the magnitude limit. The ratio $\xi/\xi^{\rm
  SAM}$ is calculated by averaging the measurements for each 10 shuffled sample on scales of
$5-25\hMpc$. We note that the trend of the ratio with the threshold luminosity is the same as in
Figure~$2$ of \cite{Croton07}, although we use a different indicator for the large scale
difference.
 
For a better understanding of the change of clustering strength in the shuffled samples with
respect to the original samples, we decompose the galaxy 2PCFs into five components according to
the source of galaxy pairs and examine them individually. The five components are denoted as
\texttt{1h-cen-sat}, \texttt{1h-sat-sat}, \texttt{2h-cen-cen}, \texttt{2h-cen-sat}, and
\texttt{2h-sat-sat}, where \texttt{1h} and \texttt{2h} refer to one-halo and two-halo pairs and
\texttt{cen} and \texttt{sat} tell the nature (central or satellite galaxies) of the pair of
galaxies. That is, we have central galaxy in a halo paired with satellites in the same halo
(\texttt{1h-cen-sat}), satellite galaxy pairs inside halos (\texttt{1h-sat-sat}), central galaxy
in one halo paired with central galaxy in a different halo (\texttt{2h-cen-cen}), central galaxy
in one halo paired with satellites in a different halo (\texttt{2h-cen-sat}), and satellites in
one halo paired with satellites in a different halo (\texttt{2h-sat-sat}). A detailed description
on how we separate these components can be found in the Appendix.

\begin{figure*}
  \epsscale{1.0} \plotone{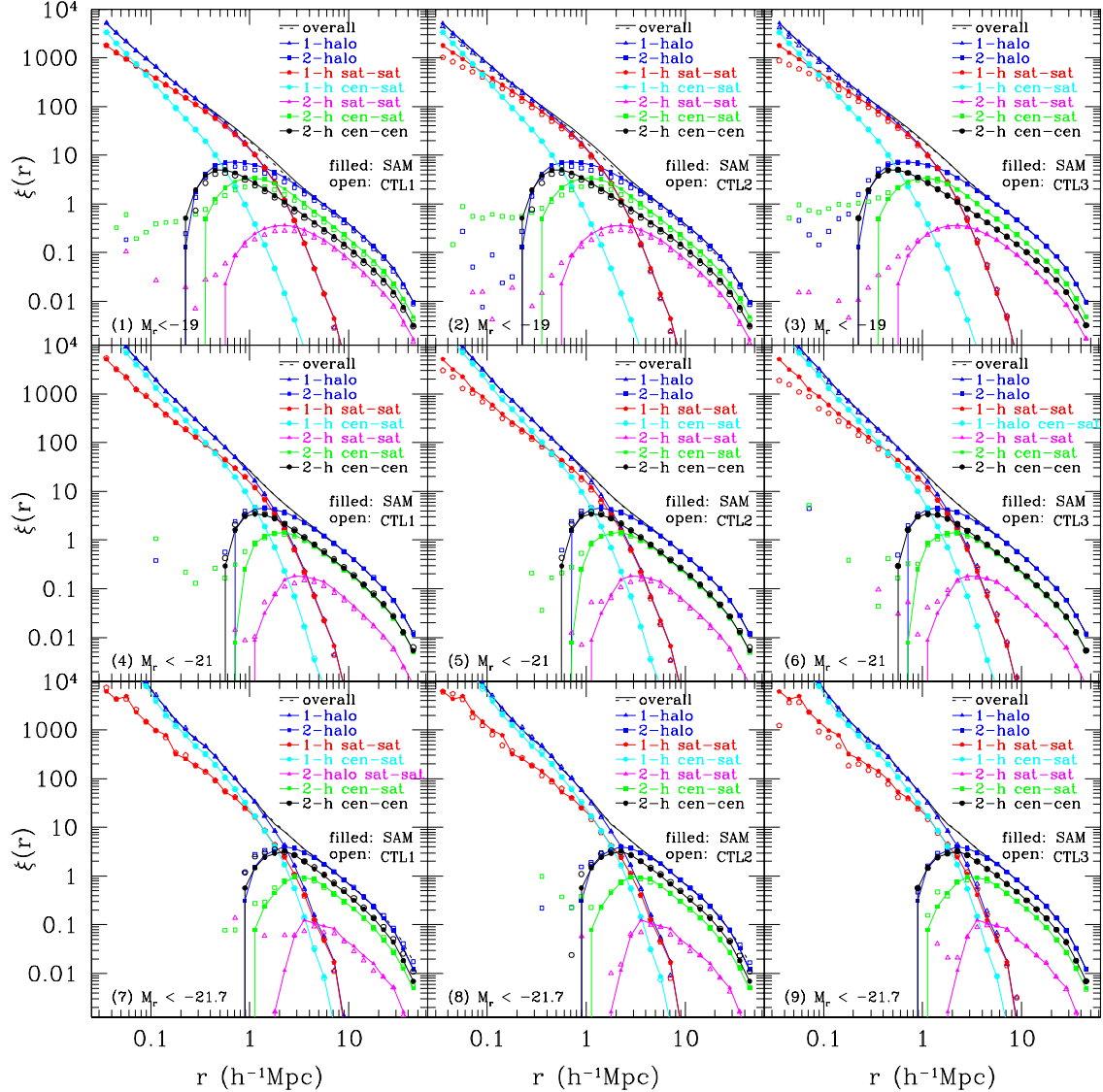}
  \caption{Comparison of 2PCFs between SAM and shuffled $M_r<-19$, -21 and -21.7 samples by
    decomposing the 2PCF into five separate components. Solid and dotted lines are the overall
    2PCF for SAM and shuffled samples, respectively. The shuffled sample in the left (middle,
    right) column is from the CTL1 (CTL2, CTL3) shuffling scheme (see the text). The five
    components correspond to contributions from one-halo central-satellite galaxy pairs, one-halo
    satellite-satellite galaxy pairs, two-halo central-central galaxy pairs, two-halo
    central-satellite galaxy pairs, and two-halo satellite-satellite galaxy pairs, respectively.
    See the text and the Appendix for more details.}
  \label{12h}
\end{figure*}

Figure~\ref{12h} shows the five 2PCF components of the original and shuffled samples for L190,
L210 and L217 (left, middle and right columns for CTL1, CTL2 and CTL3, respectively). Since the
spatial distribution of satellites inside halos is conserved in the CTL1 sample [except for the
$(M_{\rm new}/M_{\rm old})^{1/3}$ scaling], there is almost no change in the one-halo components
for this sample with respect to the original sample. In the shuffling schemes of CTL2 and CTL3
samples, satellites inside halos are angularly redistributed from a non-spherical distribution to
an isotropic distribution. The redistribution in either CTL2 or CTL3 does not change the
separations of one-halo central-satellite galaxy pairs, so the \texttt{1h-cen-sat} component does
not change after shuffling, as seen in Figure~\ref{12h}. However, the isotropization statistically
increases the separations of one-halo satellite-satellite pairs and thus dilutes the
\texttt{1h-sat-sat} clustering signal. This leads to a suppression of the 2PCF on small scales
with respect to the original sample (e.g., $\sim$10\% for L190). CTL2 samples show a smaller
suppression in the \texttt{1h-sat-sat} component than CTL3 samples. There may be two reasons for
the difference. First, CTL2 effectively uses a mean radial distribution profile of satellites in
halos of a given mass, while CTL3 uses the radial distribution in each individual halo. Because of
the scatter in the radial profiles at a given halo mass, the distributions of one-halo
satellite-satellite pair separations are not identical from the mean and individual profiles.
Second, CTL2 ensures that the numbers of satellites inside halos of a given mass follow the
Poisson distribution, while CTL3 follows the distribution in the SAM sample, which can be slightly
sub-Poisson in the low occupation regime (e.g., \citealt{Zheng05}).

For the shuffled samples CTL1, CTL2 and CTL3, all two-halo components (except \texttt{2h-cen-cen})
show enhancements on scales less than $1\hMpc$ with respect to the original ones. This is mainly
caused by the non-spherical distribution of satellite galaxies inside halos in the SAM sample. The
shuffling procedure can cause the satellite populations of two neighboring (non-spherical) halos
to become spatially close or even overlapped to some extent. Therefore, in the shuffled samples
the probability of finding close inter-halo galaxy pairs that involve satellites increases.
However, such small-scale enhancements in the two-halo components only occur on scales that the
one-halo term of the 2PCF dominates, thus, they are of no interest in our analysis.

On large scales, where the two-halo term regime dominates, the two-halo components for CTL3 do not
change since it only shuffles galaxies within halos, while every two-halo component changes its
amplitude after shuffling with CTL2 and CTL3. There is no doubt that this should be a
manifestation of the environmental dependence of the halo clustering and that of the galaxy
content inside halos. Let us first consider the \texttt{2h-cen-cen} component. The effect of
shuffling central galaxies is equivalent to that of shuffling halos. If the galaxy sample were a
halo-mass-threshold sample, shuffling would not change the large scale clustering of central
galaxies as the host halo population remains the same after shuffling. However, the sample we
consider is defined by a threshold in luminosity and it is not a halo-mass-threshold sample
because of the scatter between halo mass and central galaxy luminosity. At a fixed mass, older
halos tend to host more luminous central galaxies, and the mean central galaxy luminosity is an
increasing function of halo mass \citep{Zhu06}. We thus expect that, at a given luminosity, a
central galaxy can reside in a low mass older halo or in a younger halo of higher mass. That is,
for low mass halos, only a fraction of them (some older ones) can host the galaxies in our
luminosity-threshold sample. Since the shuffling is among halos of the same mass, some central
galaxies of the sample in these low mass older halos are moved to younger halos in the same mass
bin after shuffling. For the L190 samples, these low mass halos are in the regime where the
clustering of younger halos are weaker, so we see a decrease in the \texttt{2h-cen-cen} component
of the 2PCF after shuffling [Fig.~\ref{12h}(1) and Fig.~\ref{12h}(2)]. However, halos at the low
mass end in L217 samples are in the regime where the clustering of older halos are weaker, leading
to an increase in the \texttt{2h-cen-cen} component after shuffling [Fig.~\ref{12h}(7)
Fig.~\ref{12h}(8)]. For the L210 samples, the low mass halos are in the regime where the age
dependence of halo clustering almost disappear, and as a consequence, the \texttt{2h-cen-cen}
component does not change much after shuffling [Fig.~\ref{12h}(4) and Fig.~\ref{12h}(5)].

\begin{figure}
  \epsscale{1.} \plotone{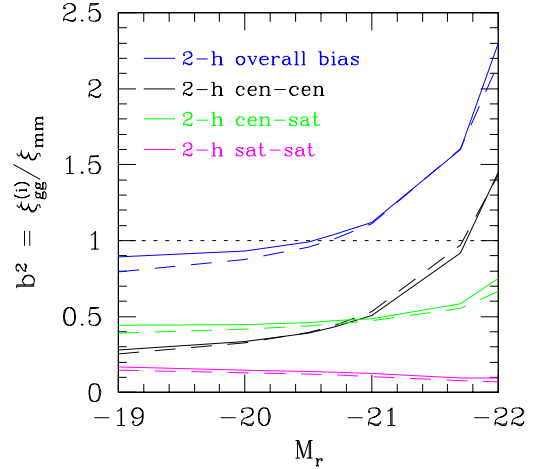}
  \caption{Contributions to the large scale bias factor from different pair components as a
    function of magnitude limit of galaxy sample. The component contribution $\xi_{gg}^{(i)}$ to
    the large scale 2PCFs, normalized by the matter 2PCF, includes \texttt{2h-cen-cen} (black),
    \texttt{2h-cen-sat} (green), and \texttt{2h-sat-sat} (magenta). The blue curve is the overall
    squared bias factor. Solid and dashed lines are for the original SAM and the CTL shuffled
    samples, respectively. See the text for more details.}
  \label{bdec}
\end{figure}

Unlike the \texttt{2h-cen-cen} component, for which the effect of shuffling is determined by the
halos near the low-mass end for the given sample, the \texttt{2h-cen-sat} and \texttt{2h-sat-sat}
components are influenced by all halos above the low-mass end. \cite{Zhu06} find that, in general,
at a fixed halo mass, there are fewer satellite galaxies in older halos. Combining this finding
with the age dependence of halo clustering (i.e., older halos being more strongly clustered in the
mass range appropriate for our sample), one would infer that shuffling would increase the
amplitudes of the \texttt{2h-cen-sat} and \texttt{2h-sat-sat} components, since the overall effect
of shuffling is to homogenize satellite populations among halos of different ages (i.e., to
increases/decrease the number of satellites in older/younger halos). However, this naive
expectation is contradictory to what is seen in Figure~\ref{12h}. Then, what is the reason for the
suppression in the \texttt{2h-cen-sat} and \texttt{2h-sat-sat} components? The answer lies in the
richness dependence of halo clustering. Here, the term ``richness'' refers to
subhalo/substructure/satellite abundance in a halo. \citet{GW07} show that, in the mass range
relevant here, halos with more substructures are always more strongly clustered. Since
substructures are the natural dwellings of satellite galaxies, we expect that, at a fixed mass,
halos that have more satellites are more strongly clustered. The effect of shuffling is to move
some satellites in strongly clustered halos to weakly clustered halos and thus lower the amplitude
of the \texttt{2h-cen-sat} and \texttt{2h-sat-sat} components of the 2PCF.

The above explanation of the two-halo term change still leaves with one question. According to the
age-dependence of halo clustering (older halos are more strongly clustered) and the
anti-correlation between age and subhalo abundance (older halos have fewer subhalos;
\citealt{Gao04,Zhu06}), one would expect that halos with fewer satellites are more strongly
clustered, in sharp contrast with what is found in simulation (e.g., \citealt{GW07}). The solution
to the apparent contradiction lies in the scatter in the anti-correlation between age and richness
and the joint dependence of halo clustering on age and richness (Zu et al. in preparation).

Figure~\ref{bdec} summarizes the contributions of two-halo components to the large-scale 2PCFs and
the changes caused by CTL1/CTL2 shuffling as a function of threshold luminosity. We plot the
contributions from different two-halo components to the large-scale bias factor (squared) for both
the original SAM samples (solid) and CTL1 samples (dashed). Each component contribution to the
square bias factor is computed by averaging the ratio of the corresponding two-halo 2PCF component
(\texttt{2h-cen-cen}, \texttt{2h-cen-sat}, or \texttt{2h-sat-sat}) to the matter 2PCF on scales of
5--15$\hMpc$. For galaxy samples with low threshold luminosity, the largest contribution to the
large-scale clustering comes from the \texttt{2h-cen-sat} component. The \texttt{2h-cen-cen}
component then takes over for samples with threshold luminosity around $L_*$ and becomes more and
more dominant towards higher luminosity. This trend can be understood by noticing that the
satellite fraction decreases with increasing threshold luminosity (e.g., \citealt{Zheng07b}). The
\texttt{2h-sat-sat} component always has the least contribution to the the large-scale clustering.

Figure~\ref{bdec} shows that shuffling causes the \texttt{2h-cen-cen} component to be suppressed
slightly for samples with low luminosity thresholds and to be enhanced a little bit for samples
with high luminosity thresholds, a trend can be explained by the age dependence of halo clustering
as discussed above. Here ``low'' and ``high'' are with respect to $L_*$. We note that the change
in the \texttt{2h-cen-cen} component decreases again at the very high luminosity end (i.e., the
L220 sample), similar to the trend seen in the concentration dependence of massive halo clustering
(see e.g., \citealt{JSM07}). The \texttt{2h-cen-sat} component is always suppressed after
shuffling, which can be understood by the richness dependence of halo clustering as mentioned
above. The change of the overall large-scale bias factor is dominated by that of the
\texttt{2h-cen-sat} at low luminosity and very high luminosity. The change of the
\texttt{2h-cen-cen} component plays a role in determining that of the overall bias factor for
samples with luminosity threshold larger but not much larger than $L_*$, and it nearly compensates
the suppression caused by the change of the \texttt{2h-cen-sat} component, leading to little
change ($<$2\%) in the overall bias factor (also see Fig.~\ref{b2}).

\section{Environmental Effect on Redshift-Space 2PCFs}

\begin{figure}
  \epsscale{1.1} \plotone{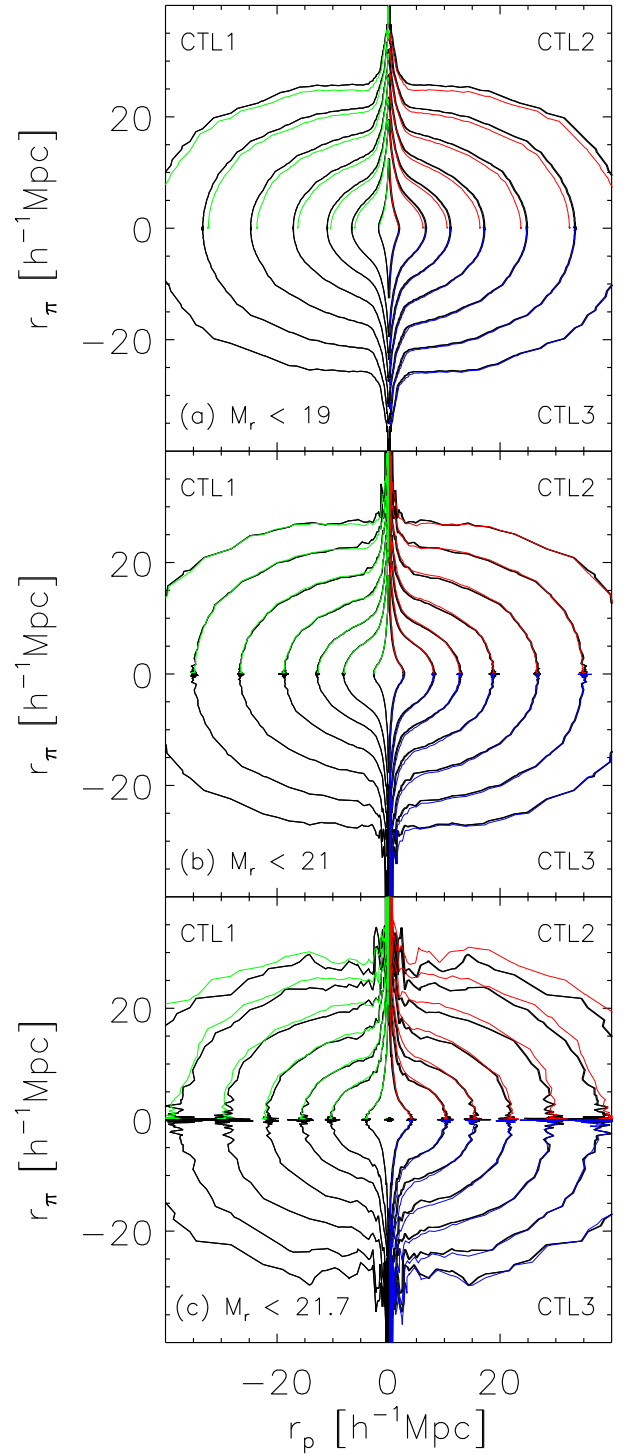}
  \caption{Comparison of redshift-space correlation functions between SAM samples and shuffled
    samples. Panels ($a$), ($b$), ($c$) are for L190, L210, and L217 samples, respectively. In
    each panel, three quadrants shows the comparison between one shuffled sample and the SAM, with
    black contours for the original SAM sample, green, red, and blue contours for the CTL1, CTL2,
    and CTL3 shuffled samples, respectively. Contour levels are set as $2^{n}$, with $n$ from -5
    to 2. }
  \label{ctr}
\end{figure}

While the 2PCFs in real space are isotropic, the 2PCFs in redshift space are distorted by galaxy
peculiar velocities along the line of sight. On small scales, the random virialized motions of
galaxies in groups and clusters stretch the redshift distribution of galaxies along the line of
sight, producing the so-called ``finger-of-god''~(FOG) effect. On large scales, the coherent flows
of galaxies due to gravity squash the line-of-sight distribution of galaxies (i.e., Kaiser effect;
\citealt{Kaiser87}).

In linear theory, the large scale redshift-space distortion measures a combination of $\Omega_m$
and the large scale galaxy bias factor $b_g$, which is $\Omega_m^{0.6}/b_g$. Given the measured
amplitude of galaxy 2PCF, a constraint on $\Omega_m^{0.6}/b_g$ is equivalent to that on
$\sigma_8\Omega_m^{0.6}$, where $\sigma_8$ is the rms matter fluctuation on scale of 8$\hMpc$. To
infer such a constraint on large scales, the small-scale redshift distortion is usually dealt with
simple models, such as the exponential model (\citealt{CFW95}). \cite{Tinker06} demonstrates that
by taking advantage of the power of HOD to describe clustering in a fully non-linear manner, one
can consistently model the small-scale and the large-scale clustering (also see
\citealt{Tinker07}). Furthermore, \cite{Tinker06} shows that the degeneracy in $\Omega_m$ and
$\sigma_8$ from large scale clustering can be broken by making use of the small- and
intermediate-scale clustering in redshift space. The HOD framework used in \cite{Tinker06} assumes
no environmental effects on halo clustering and galaxy content inside halos. In
\S~\ref{sec:realspace}, we have shown that in the SAM we use, the real-space 2PCFs may suffer a
change up to $\sim$10\% because of the environmental effect. It is interesting to perform similar
analysis in redshift space and discuss the implications in inferring cosmological parameters from
redshift distortions.

In Figure~\ref{ctr}, we show the redshift-space 2PCFs $\xi(r_p,r_\pi)$\footnote{We use the same
  symbol $\xi$ for both the real-space and redshift-space 2PCFs. Whenever it introduces a
  confusion, we will add a subscript $R$ for real-space quantities.} measured from the SAM, CTL1,
CTL2, and CTL3 catalogs for the three luminosity-threshold samples as in Figure~\ref{bias}, where
$r_p$ and $r_\pi$ are the perpendicular and line-of-sight distances in redshift space. The overall
effect of the shuffling on the redshift-space 2PCFs is similar to that seen in the real-space
2PCFs. On small scales, where the FOG effect dominates, the clustering amplitudes of CTL1 samples
(green contours) are almost identical to those of the SAM samples (black contours), since
shuffling does not change the one-halo term in CTL1. For CTL2 samples (red contours), the FOG
effect is slightly suppressed and the small scale clustering is weaker than that of the SAM for
L190, but nearly identical to and stronger than that of the SAM for L210 and L217, respectively.
For CTL3 samples (blue contours), although they exhibit large difference in real-space 2PCFs, the
FOF effect only changes a little, which indicates that the difference caused by galaxy angular
distribution is partly masked by the peculiar velocity field on small scales.

On large scales, similar to what is seen in the real-space 2PCFs, at a given large-scales
separation $(r_p,r_\pi)$, the redshift-space 2PCF of CTL1 or CTL2 sample has a lower amplitude
than that of the SAM sample for L190; it has an almost identical amplitude to that of the L210 SAM
sample; it exceeds that of the L217 SAM sample. The 2PCF amplitudes do not change with CTL3
samples. As a whole, shuffling introduces changes similar to those in the real-space 2PCFs, and
these changes can be understood following interpretations in \S~\ref{sec:realspace}.

To further quantify changes in the redshift-space 2PCFs, we calculate a few statistics derived
from the multipoles of the real-space and redshift-space 2PCFs, which were originally proposed by
\cite{Hamilton92}. These statistics are also the ones used in the study of \cite{Tinker06} and
\cite{Tinker07} for HOD modeling the redshift-space distortion.

The multipole moments $\xi_l(r)$ are given by the coefficients of the Legendre polynomial
expansion of $\xi(r_p,r_{\pi})$.
\begin{equation}
\label{equ-1}
\xi_l(r)=\frac{2l+1}{2}\int_{-1}^{+1}\xi(r_p,r_{\pi})P_l(\mu)d\mu
\end{equation}
where $r=\sqrt{r^2_p+r^2_{\pi}}$, $\mu=r_{\pi}/r$, and $P_l(\mu)$ is the $l$-th order Legendre
polynomial. Based on the multipoles of $\xi(r_p,r_{\pi})$, we calculate two statistics. The first
one is the ratio of the monopole $\xi_0(r)$ to the real-space 2PCF $\xi_R(r)$,
\begin{equation}
\label{equ-2}
\xizr(r) \equiv \frac{\xi_0(r)}{\xi_R(r)}. 
\end{equation}
In linear theory, it is a function of $\beta\equiv\Omega_m^{0.6}/b_g$,
\begin{equation}
\label{eqn:mono_lin}
\xizr(r)=1 + \frac{2}{3}\beta + \frac{1}{5}\beta^2.
\end{equation}
The second quantity $\xiQp(r)$ is related to the quadrupole $\xi_2(r)$,
\begin{equation}
\label{equ-3}
\xiQp(r) \equiv \frac{\xi_2(r)}{\xi_0(r) - \bar{\xi}_0(r)}, 
\end{equation}
where $\bar{\xi}_0(r)$ is the volume-averaged monopole,
\begin{equation}
\label{equ-4}
\bar{\xi}_0(r) = \frac{3}{r^3}\int_0^r\xi_0(s) s^2 ds.
\end{equation}
In linear theory, $\xiQp(r)$ is also a function of $\beta$,
\begin{equation}
\label{eqn:quad_lin}
\xiQp(r)=\frac{\frac{4}{3}\beta + \frac{4}{7}\beta^2}{1+\frac{2}{3}\beta + \frac{1}{5}\beta^2}.
\end{equation}
\cite{Tinker06} also introduce a quantity $\rhalf$, which is the value of $r_\pi$ at which the
redshift-space 2PCF at the given $r_p$ decreases by a factor of 2 with respect to the value
of 2PCF at $r_\pi=0$. We also compute this quantity.
\begin{figure*}
  \epsscale{1.} \plotone{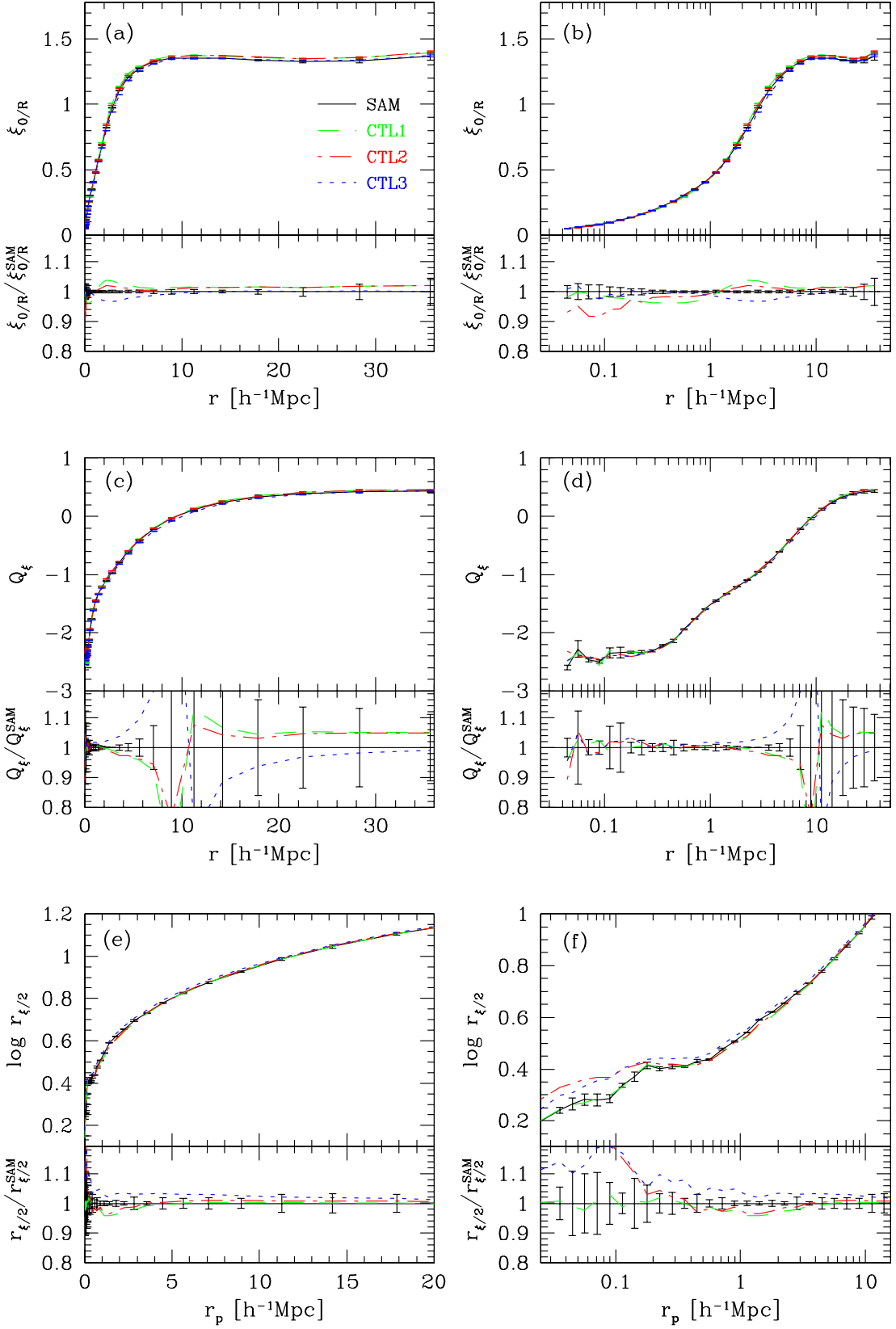}
  \caption{Comparison of redshift-space clustering statistics between the SAM and shuffled samples
    for the L190 galaxy sample. Top panels are for $\xizr$, the ratio of the monopole of the
    redshift-space 2PCF to the real-space 2PCF; middle panels are for $\xiQp$, which is related to
    the quadrupole of the redshift-space 2PCF; bottom panels are for $\rhalf$, which is the value
    of $r_\pi$ at which the redshift-space 2PCF at a give $r_p$ decreases by a factor of 2
    relative to its value at $r_\pi=0$. In each panel, black solid curves stand for the SAM
    sample, while the green dashed, the red dot-dashed and the blue dotted curves indicate the
    CTL1, CTL2, and CTL3 shuffled samples, respectively. The horizontal axes are shown in linear
    (logarithmic) scales in the left (right) panels to highlight the large (small) scale behavior.
    Error bars are plotted only for the SAM sample to avoid crowding and those for the shuffled
    ones are comparable. The abnormal error bars around 10 $\hMpc$ in the middle panels of
    panels~($c$) and~($d$) are because of $\xiQp$ there approaching zero. }
  \label{mut}
\end{figure*}

Figure~\ref{mut} plots $\xizr$ and $\xiQp$ as a function of $r$ and $\rhalf$ as a function of
$r_p$ for the L190 sample using both linear and logarithmic axes to highlight large and small
scales separately. According to the above results, this sample, among the six luminosity-threshold
samples, is expected to show the largest environmental effect (we also check the
luminosity-threshold sample with magnitude limit $M_{r}=-18.0$ and find that the large scale
suppression is still at the 10\% level as it is in L190).

Compared to the SAM sample, the monopole term $\xizr$ in either CTL1 or CTL2 is only $\sim$2\%
higher on large scales, where it stays the same in CTL3. On small scales (0.1--1$\hMpc$), the
difference is at a level of 5\% in CTL1\&2. Only on extremely small scales ($\lesssim 0.1\hMpc$)
does $\xizr$ of the CTL2 sample show a 10\% drop, which is not important since the error bars are
large and these scales are likely to be excluded in HOD modeling. In the CTL3 sample, since the
suppression of the spherically-averaged redshift-space 2PCFs and that of the real-space 2PCFs
cancel with each other on small scales, $\xizr$ stays at the same level as in the SAM sample.

For the quadrupole term $\xiQp$, the difference between the results of the original and shuffled
samples is well within 5.5\% in CTL1/CTL2 for most scales. Note that the difference extends all
the way to the largest scales in CTL3 at a 0.5\% level, which means that the non-linearity still
affects clustering behaviors on linear scales in redshift-space. Also note that the large
fractional differences around 10$\hMpc$ are simply because $\xiQp$ is crossing zero. The
fractional changes in $\xizr$ and $\xiQp$ caused by shuffling are much less than those in the
real-space 2PCFs, which are at a level of 10\%. For the quantity $\rhalf$, the global enhancement
in CTL3 is caused only by the angular isotropizing of galaxies, which disrupts the original
compact configuration of SAM halos. This makes the redshift-space 2PCFs harder to decrease to the
half value of $\xi|_{r_\pi=0}$ at a given $r_{p}$, especially at $r_p<0.2\hMpc$, where the
enhancement becomes much eminent. In CTL2, $\rhalf$ shows a similar behavior as in CTL3 at small
$r_{p}$ for the same reason, then it becomes smaller than that in CTL3 at $r_p>0.2\hMpc$, while
$\rhalf$ from the CTL1 shuffled sample is consistent with that from the original sample within the
error bars.

How would the above changes in the redshift distortion statistics induced by the environmental
effect affect the inference of cosmological parameters from HOD modeling? For a complete answer of
this question, one needs to perform the analysis presented in \citet{Tinker07} with the 2PCF
measurements in the original and shuffled samples and compare the results in the inferred
cosmological parameters. However, even without the full analysis, we can still figure out the
likely magnitude by using the linear theory results [equations~(\ref{eqn:mono_lin}) and
(\ref{eqn:quad_lin})]. For the L190 sample presented in Figure~\ref{mut}, a 2\% increase in the
large-scale $\xizr$ only leads to $\sim$6.5\% increase in the inferred $\beta$. For $\xiQp$,
shuffling gives rise to a 5.5\% increase on large scales, which also translates to a $\sim$6.5\%
increase in $\beta$. For $\rhalf$, it is not straightforward to see the consequence. Based on
Figure 16 of \citet{Tinker06}, it is likely that the difference in $\rhalf$ on large scales
between the SAM and shuffle samples can at most lead to a $\sim$5\% uncertainty in constraining
$\sigma_8$. We note that the effects of the shuffling on both the real-space and redshift-space
clustering statistics are likely from the same cause in that the $\sim $12\% decrease in the
real-space 2PCF $\xi_{R}(r)$ leads to 6\% decrease in galaxy bias $b_{g}$, which in turn
corresponds to a $\sim$6.5\% increase in $\beta$, about the number we infer from redshift-space
clustering statistics. Since the shuffling induced changes in the real-space 2PCFs for other
brighter samples are smaller, the environmental effect on $\beta$ is expected to be smaller for
them. Therefore, in the SAM galaxy catalog we use, neglecting any environmental dependence of halo
clustering and galaxy formation is likely to cause a less than 7\% systematic uncertainty in
constraining $\sigma_8\Omega_m^{0.6}$.

\section{Summary and Discussion}

In this work, we investigate the effect of environmental dependence of halo clustering and galaxy
formation on real-space and redshift-space clustering of galaxies. Our study makes use of the
galaxy catalog from the SAM of \cite{DB07}, which is based on the Millennium Simulation
\citep{Springel05}. The inherent dependence of galaxy properties on environment in the SAM catalog
is eliminated by shuffling galaxies among halos of similar mass.

The real-space 2PCFs in the original sample and those in the shuffled samples have a difference at
the level of 10\% with some dependencies on scales for samples with low threshold luminosities.
The difference becomes much smaller for samples with threshold luminosities approaching or
exceeding $L_*$. We decompose the 2PCFs into five components by accounting for the nature of
galaxy pairs (e.g., one-halo or two-halo, central galaxies or satellites) and study the effect of
environment on each of them. In general, on large scales, the changes in the \texttt{2h-cen-cen}
component of the 2PCF caused by shuffling can be well understood by noticing the dependence of
halo bias and central galaxy luminosity on halo formation time, while those in the
\texttt{2h-cen-sat} and \texttt{2h-sat-sat} components are determined by the richness
(substructure) dependence of halo bias. The \texttt{2h-cen-sat} component appears to dominate the
change in the overall 2PCF for samples with low or very high threshold luminosity, while the
change in the \texttt{2h-cen-cen} component nearly compensates that in the \texttt{2h-cen-sat}
component for threshold luminosity $L_*$, where the amplitude of the environment effect is small.
These results imply that we could use high-resolution $N$-body simulations to accurately model the
2PCFs by associating satellites to substructures identified in halos and neglecting the
environmental dependence of central galaxies at fixed halo mass. On small scales, the assumption
of spherical symmetry in galaxy distribution may lead to an uncertainty as large as 10\% in 2PCFs,
but this effect could be absorbed into a free parameter describing the halo concentration.
 
The effects of environmental dependence on redshift-space 2PCFs are similar to what are seen in
the real-space 2PCFs. On large scales, the effects can be attributed solely to the change in the
large scale bias factor. For inferring cosmological parameters ($\sigma_8$ and $\Omega_m$) through
HOD modeling of the redshift-space distortion (\citealt{Tinker06,Tinker07}), the systematic effect
caused by neglecting the environmental dependence of halo clustering and galaxy formation is
likely to be at the level of $<$6.5\% for the worst case (the $M_r<-19.0$ or $L>0.2L_*$ sample)
and can be much smaller for brighter samples, especially for samples with threshold luminosities
near $L_*$. The underlying assumption of this statement is that the environmental effect on galaxy
formation in reality is as large as that seen in the SAM we use in this paper.

Our results are based on one particular galaxy formation model, i.e., SAM of \cite{DB07}. Although
this model can reproduce many observed properties of galaxies, it is not guaranteed to be
absolutely correct. In this model, the environmental effect on the formation and evolution of
galaxies is mostly linked to the formation/merger history of dark matter halos. Compared to
observations, it overproduces faint red galaxies \citep{Croton06}. Although the model predicts the
correct trend of the color dependent galaxy clustering, it predicts too large a difference between
the amplitudes of the 2PCFs of blue and red galaxies \citep{Springel05}. By comparing to the
galaxies in SDSS groups, \cite{Weinmann06} find that SAM produces too many faint satellites in
massive halos and incorrect blue fractions of central and satellites galaxies.

The discrepancies between observations and the SAM model suggest that the effect of environment on
galaxy formation and evolution may be exaggerated in this particular model. Such discrepancies
provide opportunities for enhancing our understanding of galaxy formation and evolution. There are
also tests with void statistics \citep{Tinker07b}, environmental dependence of group galaxies
\citep{Blanton07}, and marked galaxy correlation function \citep{Skibba06}, which show that the
observed properties of galaxies are mainly driven by host halo mass rather than the environment in
which halos form. Therefore, in reality, it is quite possible that the environmental effect on
modeling galaxy clustering statistics is much smaller than what we obtain in this paper, and that
the systematic effect on cosmological parameter constraints from HOD modeling is not larger than a
few percent or even better.

\acknowledgments 

We thank Jeremy Tinker for helpful discussions and David Weinberg for useful comments. Y. Z. and
Y. P. J. are supported by NSFC (10533030), by the Knowledge Innovation Program of CAS (No.
KJCX2-YW-T05), by 973 Program (No.2007CB815402), and by Shanghai Key Projects in Basic research
(05XD14019). At an earlier stage of this work, Z. Z. was supported by NASA through Hubble
Fellowship grant HF-01181.01-A awarded by the Space Telescope Science Institute, which is operated
by the Association of Universities for Research in Astronomy, Inc., for NASA, under contract NAS
5-26555. Z. Z. gratefully acknowledges support from the Institute for Advanced Study through a
John Bahcall Fellowship. The Millennium Simulation databases used in this paper and the web
application providing online access to them were constructed as part of the activities of the
German Astrophysical Virtual Observatory.

\newpage

\appendix
\section{Decomposition of the Two-Point Correlation Function}

In the HOD framework, the 2PCF $\xi(r)$ is usually decomposed into two components (e.g., 
\citealt{Zheng04}),
\begin{equation}
  \label{app0}
  \xi(r) = [1 + \xi_{\rm 1h}(r)] + \xi_{\rm 2h}(r),
\end{equation}
where the one-halo term $\xi_{\rm 1h}(r)$ and the two-halo term $\xi_{\rm 2h}(r)$ represent
contributions from intra-halo and inter-halo pairs, respectively. To separate such components from
measurements in a mock catalog, one only needs to weigh the total correlation function
appropriately. That is, $1+\xi(r)$ weighted by the fraction of intra-halo (inter-halo) pairs at a
separation $r$ gives $1+\xi_{\rm 1h}(r)$ [$1+\xi_{\rm 2h}(r)$]. However, the way to decompose
$\xi(r)$ into more components on the basis of pair counts, like what we do in this paper
(central/satellite, one-halo, two-halo pairs), is not immediately clear. In this Appendix, we
develop a method for such a decomposition. The method can be generalized to apply to real data:
for example, one is able to measure the two-point auto-correlation functions of red and blue
galaxies and their two-point cross-correlation functions in a single run with only one random
catalog.

We first provide a general consideration on the component separation and then describe the
decomposition used in this paper in more details.

\subsection{General Consideration}

\label{a1}

Let us start from the definition of the two-point correlation function,
\begin{equation}
\label{app1}
 \xi(\rvec)=\la \delta(\xvec)\delta(\xvec+\rvec)\ra,
\end{equation}
where $\la...\ra$ represents an ensemble average. The overdensity field $\delta$ is defined as
\begin{equation}
\label{app2}
 \delta(\xvec)=\frac{n(\xvec)-\bar{n}}{\bar{n}},
\end{equation}
where $n(\xvec)$ is the galaxy density at $\xvec$ and $\bar{n}$ is the mean. Let us assume that
the galaxy sample is composed of several sub-samples, $n(\xvec)=\sum_i n_i(\xvec)$. Now we
decompose the overdensity into different components based on sub-samples

\begin{equation}
  \label{app3}
 \delta=\sum_i \tilde{\delta}_i,
\end{equation}
where $\tilde{\delta}_i$ is the overdensity contributed by the $i$-th component (sub-sample),
\begin{equation}
\label{app4}
\tilde{\delta}_i=\frac{n_i-\bar{n}_i}{\bar{n}}=\frac{n_i-\bar{n}_i}{\bar{n}_i} \times \frac{\bar{n}_i}{\bar{n}} = \delta_i \frac{\bar{n}_i}{\bar{n}}.
\end{equation}
Note that in the above equation, $\delta_i$ is the $i$-th component's own overdensity field (i.e.,
fractional fluctuation with respect to $\bar{n}_i$ instead of $\bar{n}$).

Substituting equations (\ref{app3}) and (\ref{app4}) into (\ref{app1}), we obtain
\begin{equation}
\label{app5}
\xi(\rvec)= \sum_i \la\delta_i(\xvec)\delta_i(\xvec+\rvec)\ra 
\frac{{\bar{n}_i}^2}{\bar{n}^2}
+ \sum_{i<j} \la\delta_i(\xvec)\delta_j(\xvec+\rvec)\ra
\frac{2\bar{n}_i\bar{n}_j}{\bar{n}^2}.
\end{equation}
That is, the total correlation function is a weighted sum of the auto- and cross-correlation
functions of all components, where the weight is the pair fraction. In terms of measurement 
from pair counts in a galaxy catalog and an auxiliary random catalog, it converts to
\begin{equation}
\label{app6}
\xi(\rvec) = \sum_{i\leq j} \frac{dd_{ij}(\rvec)-rr_{ij}(\rvec)}{rr_{ij}(\rvec)} 
f_{ij},
\end{equation}
where $dd_{ij}$ and $rr_{ij}$ are the $ij$ data-data and random-random pairs. The quantity
$f_{ij}$ is the overall $ij$ pair fraction (the ratio of the total number of {\it ij} pairs in the
volume to that of all pairs in the volume). Note that, for random pairs, the ratio of the number
of random $ij$ pairs to that of all random pairs is independent of separation, i.e.,
$f_{ij}=rr_{ij}(\rvec)/RR(\rvec)$, where $RR$ is the number of random pairs for all galaxies.
Therefore, we have the contribution from the $ij$ component as
\begin{equation}
\label{app7}
 \xi_{ij}(\rvec)=\frac{dd_{ij}(\rvec)-rr_{ij}(\rvec)}{RR(\rvec)},
\end{equation} 
where $rr_{ij}/RR$ is a known quantity given the number density of each component and one only
needs to measure $dd_{ij}(\rvec)$ and $RR(\rvec)$.

An interesting application of the above results to real data is to measure 2PCFs (either projected
ones or redshift-space ones) of sub-samples of galaxies of a volume-limited sample (e.g., a sample
of galaxies divided into blue, green, and red galaxy sub-samples). To measure all the two-point 
auto-correlation functions of galaxies in the sub-samples and their two-point cross-correlation 
functions, we do not need to construct random catalogs for each sub-sample. We only need one random 
catalog for the whole sample and measure all the correlation functions in a single run based on 
equation~(\ref{app7}). The two-point auto-correlation function for all the galaxies in the whole 
sample, as the weighted sum of the two-point auto- and cross-correlation functions of sub-samples
(eq.[\ref{app5}]), is obtained for free. To generalize equation~(\ref{app7}) in the spirit of the 
widely used Landy-Szalay estimator \citep{Landy93}, one can replace $-rr_{ij}(\rvec)$ with 
$-2dr_{ij}(\rvec)+rr_{ij}(\rvec)$. To count $dr_{ij}$ data-random pairs, one may randomly tag 
the points in the random catalog with component indices according to the fraction of the sub-sample 
galaxy spatial density in the overall sample.

\subsection{Details on the Decomposition of the 2PCF into Central/Satellite and One-halo/Two-halo Terms}

Following similar reasoning as in \S~\ref{a1}, now let us tag galaxies with two subscripts and
decompose the overdensity field as
\begin{equation}
\label{app8}
\delta=\sum_{i,\alpha} \tilde{\delta}_{i\alpha},
\end{equation}
where $i$ denotes the ID of the host halo and $\alpha$ is either $c$ (central) or $s$ (satellite).
That is, the overdensity is decomposed into contributions from central and satellite galaxies from
each halo. The random catalog can be obtained by randomly redistributing all the galaxies in the
volume with their tags untouched.

In a similar way as before, we can write $\tilde{\delta}_{i\alpha}$ as
\begin{equation}
\label{app9}
\tilde{\delta}_{i\alpha}=\frac{n_{i\alpha}-\bar{n}_{i\alpha}}{\bar{n}_{i\alpha}} \times \frac{\bar{n}_{i\alpha}}{\bar{n}} = \delta_{i\alpha} \frac{\bar{n}_{i\alpha}}{\bar{n}}.
\end{equation}
It is straightforward to show that $\xi(\rvec)$ can be formally decomposed as
\begin{eqnarray}
\label{app10}
\xi(\rvec)&=& \,\,\,\,\,
\sum_i \la \delta_{ic}(\xvec)\delta_{ic}(\xvec+\rvec) \ra 
\frac{\bar{n}_{ic}^2}{\bar{n}^2} \nonumber \\
& & +
\sum_i \la \delta_{ic}(\xvec)\delta_{is}(\xvec+\rvec) \ra 
\frac{2\bar{n}_{ic}\bar{n}_{is}}{\bar{n}^2} \nonumber \\
& & +
\sum_i \la \delta_{is}(\xvec)\delta_{is}(\xvec+\rvec) \ra 
\frac{\bar{n}_{is}^2}{\bar{n}^2} \nonumber \\
& & +
\sum_{i<j} \la \delta_{ic}(\xvec)\delta_{jc}(\xvec+\rvec) \ra 
\frac{2\bar{n}_{ic}\bar{n}_{jc}}{\bar{n}^2} \nonumber \\
& & +
\sum_{i \ne j} \la \delta_{ic}(\xvec)\delta_{js}(\xvec+\rvec) \ra
\frac{\bar{n}_{ic}\bar{n}_{js}}{\bar{n}^2} \nonumber \\
& & +
\sum_{i<j} \la \delta_{is}(\xvec)\delta_{js}(\xvec+\rvec) \ra 
\frac{2\bar{n}_{is}\bar{n}_{js}}{\bar{n}^2}.
\end{eqnarray}

It is easy to identify the six terms in the right hand side as contributions from the one-halo
cen-cen (which is a Dirac $\delta_D$ function that we are not interested in), the one-halo
cen-sat, the one-halo sat-sat, the two-halo cen-cen, the two-halo cen-sat, and the two-halo
sat-sat pairs, respectively (`cen' for central galaxy and `sat' for satellite galaxy).

In terms of measurement, each component can be reduced to the $(dd-rr)/RR$ form. As an example,
consider the case for the two-halo cen-sat term. Note that $\bar{n}_{ic}=1/V$,
$\bar{n}_{is}=N_{is}/V$, and $\bar{n}=N/V$, where $N_{is}$ is the number of satellites in the halo
of ID $i$ and $N$ is the total number of all galaxies in the volume $V$. Therefore, the two-halo
cen-sat contribution is
\begin{equation}
\label{app11}
\xi_{{\rm 2h},cs}(\rvec)=
\sum_{i \ne j} \la \delta_{ic}(\xvec)\delta_{js}(\xvec+\rvec) \ra 
\frac{\bar{n}_{ic}\bar{n}_{js}}{\bar{n}^2} 
= \sum_{i \ne j} \frac{dd_{ic,js}(\rvec)-rr_{ic,js}(\rvec)}{rr_{ic,js}(\rvec)}
\frac{N_{js}}{N^2},
\end{equation}
where $dd_{ic,js}$ and $rr_{ic,js}$ are numbers of data-data and random-random pairs between
galaxies tagged as $ic$ and $js$. One thing to notice is that $2rr_{ic,js}(\rvec)/N_{js}$ does not
depend on $i$ and $j$ --- it equals $rr_{cs^\prime}(\rvec)/N_{{\rm pair},cs^\prime}$, where
$rr_{cs^\prime}(\rvec) =\sum_{i \ne j} rr_{ic,js}(\rvec)$ is the count of all the random
``two-halo'' cen-sat pairs with separation around $\rvec$ and $N_{{\rm pair},cs^\prime}$ is the
total number of ``two-halo'' cen-sat pairs in the volume ($cs^\prime$ denotes a ``two-halo''
pair). Also noting that $N^2/2=N_{{\rm pair},total}$ ($N \gg1$), we then have
\begin{equation}
\label{app12}
\xi_{{\rm 2h},cs}(\rvec)=\frac{dd_{cs^\prime}(\rvec)-rr_{cs^\prime}(\rvec)}{rr_{cs^\prime}(\rvec)/N_{{\rm pair},cs^\prime}}\frac{1}{N_{{\rm pair},total}},
\end{equation}
where $dd_{cs^\prime}$ and $rr_{cs^\prime}$ are all the data-data and random-random ``two-halo''
cen-sat pairs with separation around $\rvec$. We have the following relation between
$rr_{cs^\prime}$ and the total number of all random pairs around separation $\rvec$, $RR(\rvec)$,
\begin{equation}
\frac{rr_{cs^\prime}(\rvec)}{N_{{\rm pair},cs^\prime}}=\frac{RR(\rvec)}{N_{{\rm pair},total}}.
\end{equation}
Therefore, we end up with
\begin{equation}
\xi_{{\rm 2h},cs}(\rvec)
=\frac{dd_{cs^\prime}(\rvec)-rr_{cs^\prime}(\rvec)}{RR(\rvec)}.
\end{equation}



\end{document}